\documentstyle[aps,multicol,psfig]{revtex}
\input epsf
\def\be{\begin{equation}}
\def\ee{\end{equation}}
\def\bea{\begin{eqnarray}}
\def\eea{\end{eqnarray}}
\def\bml{\begin{mathletters}}
\def\eml{\end{mathletters}}
\def\SC{{$\Lambda$}}
\def\SCm{{\Lambda}}
\def\gf{{\bf G}}

\def\bdg{BdG}

\def\pvr{{\bf r}}
\def\psr{{    r}}
\def\fwv{\kappa}
\def\Kf{\fwv}
\def\Ef{\fwv^{2}}

\def\idmat{{\rm I}}
\def\sudel{{\Delta}}
\def\spp{\sudel}
\def\ipe{\Gamma}
\def\dos{\rho}

\def\qee{\varepsilon}
\def\ima{{\rm Im}}
\def\tr{{\rm Tr}}
\def\ab{AB}
\def\surS{\partial{\cal V}}     

\def\psm#1{{\bbox{\sigma}_{#1}}}        
\def\s1{{\psm{1}}}
\def\s2{{\psm{2}}}
\def\s3{{\psm{3}}}

\def\ham{{\hat{h}}}

\newcommand{\eqbreak}{
\end{multicols}
\widetext
\noindent
\rule{.48\linewidth}{.1mm}\rule{.1mm}{.1cm}
}
\newcommand{\eqresume}{
\noindent
\rule{.52\linewidth}{.0mm}\rule[-.1cm]{.1mm}{.1cm}\rule{.48\linewidth}{.1mm}
\begin{multicols}{2}
\narrowtext
}
\begin{document}
\draft
\title{Quantal Andreev Billiards:
        Density of States Oscillations and \\
        the Spectrum-Geometry Relationship}

\author{\.Inan\c{c} Adagideli
and
Paul M. Goldbart
}
\address{Department of Physics,
University of Illinois at Urbana-Champaign, 1110 West Green
Street, Urbana, Illinois 61801-3080
}
\maketitle
\begin{abstract}
Andreev billiards are finite, arbitrarily-shaped, normal-state
regions, surrounded by superconductor.  At energies below the
superconducting energy gap, single-quasiparticle excitations
are confined to the normal region and its vicinity, the
mechanism for confinement being Andreev reflection. Short-wave quantal
properties of these excitations, such as the connection between the
density of states and the geometrical shape of the billiard,
are addressed via a multiple scattering approach.
It is shown that one can, {\it inter alia\/}, \lq\lq hear\rq\rq\
the stationary chords of Andreev billiards.
\end{abstract}
\begin{figure}[hbt]
\epsfxsize=\columnwidth
\centerline{\epsfbox{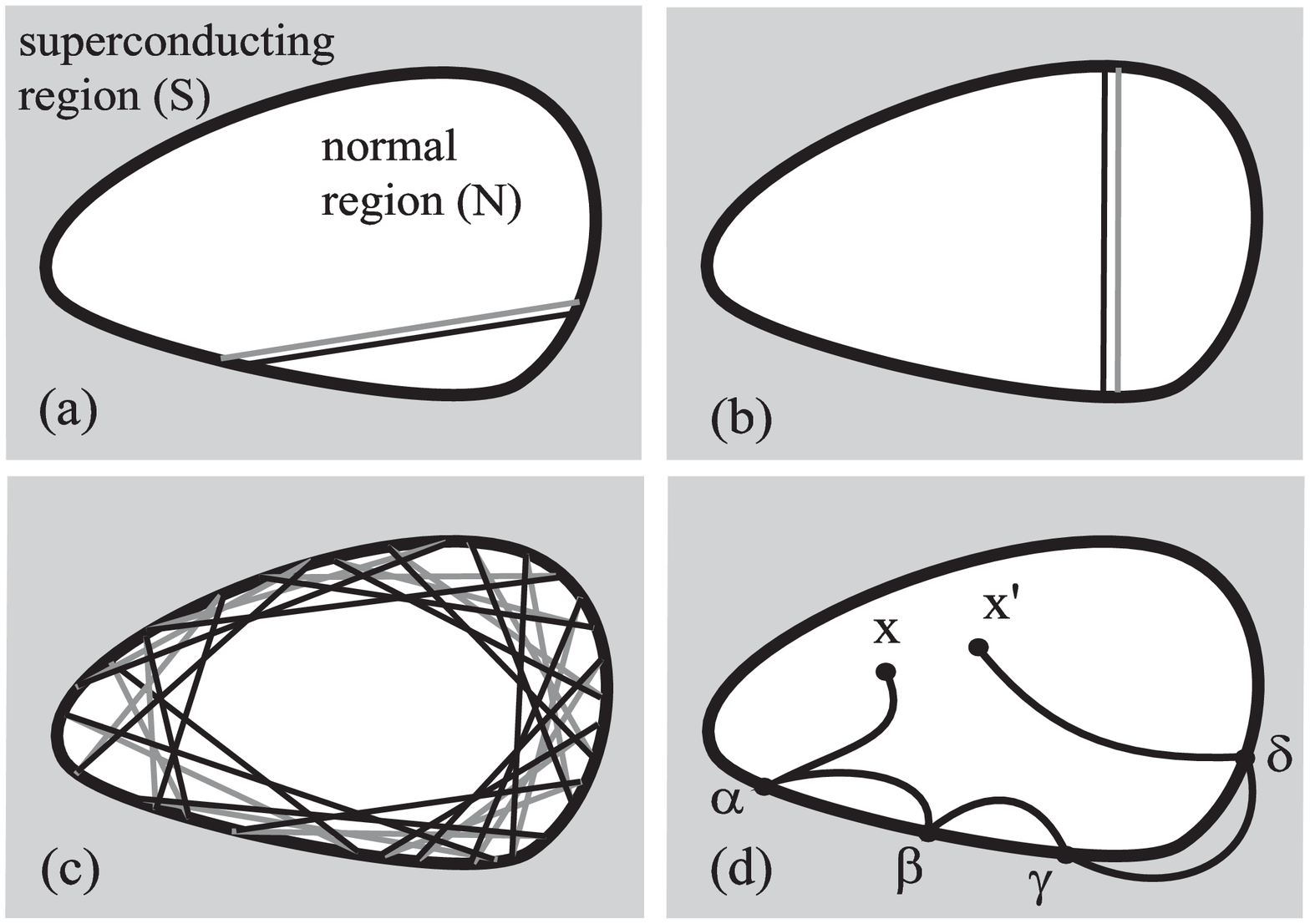}}
\vskip+0.4truecm
\caption{%
(a)~Andreev billiard, showing a generic retro-reflecting orbit.
(b)~Orbit corresponding to a stationary chord.  The length
of this orbit should be contrasted with that in~(c), which
shows a generic \lq\lq creeping\rq\rq\ orbit due to imperfect
retro-reflection; electrons (holes) follow full (shaded) lines.
(d)~Generic term in the multiple scattering expansion;
internal (external) lines represent
Green functions
$\gf^{\rm N}$($\gf^{\rm S}$).%
}
\label{FIG:billiard}
\end{figure}

%
%
%
\begin{multicols}{2}
\narrowtext
\noindent
{\it Introduction\/}.
The aim of this Letter is to explore certain quantal aspects of
quasiparticle motion arising in a class of mesoscopic structures
known as Andreev billiards (\ab s).
By the term Andreev billiard~\cite{REF:Kosztin1995}
we mean a connected, normal-state region (N) completely surrounded
by a conventional superconducting region~(S), as sketched for the 2D
case in Fig.~\ref{FIG:billiard}a.  The S region is responsible for
confining quasiparticles that have energies less than the
superconducting energy gap to the normal region and its
neighborhood~\cite{REF:Kosztin1995,REF:Rmatrix}.
The terminology \ab\ reflects the centrality of the role played by
quasiparticle reflection from the surrounding
pair-potential~\cite{REF:AFAndreev}.
Our focus here will be on the
the density of energy levels of the quasiparticle states localized
near a generically shaped \ab, and its relationship to the
geometrical shape of the \ab.
The main new features that our approach is able to capture are
the oscillations in the level-density caused by the spatial
confinement of the quasiparticles.  This structure is
inaccessible via conventional quasiclassical methods.
\par
Our central results concern the density of states (DOS) for
billiards of arbitrary shape and dimensionality.  They include
an explicit formula for the coarse DOS, as well as a general method
for obtaining the (previously inaccessible) oscillations about
this coarse DOS, both valid in the short-wave limit.  This DOS
decomposition is feasible because, unlike for conventional
billiards, the classical trajectories of {\ab}s fall into two
well-separated classes:
(i)~tracings of stationary chords (see Fig.~\ref{FIG:billiard}b),
and (ii)~certain extremely long trajectories with many reflections
(see Fig.~\ref{FIG:billiard}c), which contribute to the level density
only at fine energy resolutions.
\par
Our strategy for exploring the quantal properties of {\ab}s is as
follows.
First, we express the Green function for the appropriate
(i.e.~Bogoliubov-de~Gennes~\cite{REF:PGdGbook}; henceforth \bdg)
single-quasiparticle energy eigenproblem as an expansion in terms of
various scattering processes from the N-S interface.
Next, we identify which of these scattering processes dominate, by
effectively integrating out processes that involve propagation inside
the S region, thus arriving at an expansion involving only
reflections (i.e.~scattering processes that keep the quasiparticles
{\it inside\/} the billiard).
The processes associated with these reflections can be
classified as those that interconvert electrons and holes (which we
refer to as Andreev reflections, and which typically dominate)
and those that do not (which we refer to as ordinary
reflections).
We then compute the {\it oscillatory\/} part of the DOS
via two distinct asymptotic schemes.
\par
The first scheme amounts to an elaboration of that adopted by Andreev,
and is what is conventionally understood when the terms semiclassical
or quasiclassical are used in the subject of superconductivity. Its
physical content is perfect retro-reflection (i.e.~velocity reversal)
of quasiparticle excitations from the N-S boundary and perfect e/h
(i.e.~electron/hole) interconversion (i.e.~the neglect of ordinary
reflection processes).  It yields a smooth (i.e.~low energy-resolution)
DOS, as well as singular features that arise from stationary-length
chords. However, it is incapable of capturing other features in the DOS
caused by the spatial confinement of quasiparticles.
\par
The second scheme incorporates the effect of the imperfectness of
retro-reflection which results from differences between, say,
incident e and reflected h wave vectors, as well as the effect of
ordinary reflection processes.  It yields the DOS with higher
energy-resolution, thus revealing the oscillations caused by spatial
confinement. In order to distinguish the effect of imperfect e/h
interconversion from higher-order quantum effects, we introduce and
study a model that features perfect e/h interconversion but still
includes all quantal effects.  This model is also useful when the
pair-potential varies smoothly (so that ordinary reflection is even
more strongly suppressed).
\par
Finally, for the purpose of illustration we examine the the case
of a two-dimensional circular billiard, and compare the predictions
for the DOS obtained via the various asymptotic schemes with those
arising from the exact numerical treatment of the full \bdg\
eigenproblem, as well as from the perfect e/h interconverting model.
This provides a concrete illustration of the implications of wave
phenomena for the quasiparticle quantum states of \ab s.
\par\smallskip\noindent
{\it Eigenproblem for the Andreev billiard;
formulation as a boundary integral equation\/}.
To address the \bdg\ eigenproblem for \ab s we focus
on the corresponding ($2\times 2$) Green function $\gf$,
which obeys
\be
\label{EQ:defgf}
\pmatrix{-\ham+z&\spp(\pvr)\cr
\spp^{\ast}(\pvr)&\ham+z\cr}
\gf(\pvr,\pvr';z)
=-\idmat\,\delta(\pvr-\pvr'),
\ee
where
$\ham\equiv-\nabla^{2}-\fwv^{2}$, together with
the boundary condition that $\gf$ should vanish in the limit
of large $|\pvr|$.  Here, $\pvr$ and $\pvr'$ are spatial coordinates,
$\hbar^{2}\fwv^{2}/2m$ is the Fermi energy (i.e.~$\fwv$ is the Fermi
wave vector), $\hbar^{2}z/2m$ is the (complex) energy, and
$\hbar^{2}\spp(\pvr)/2m$ is the position-dependent superconducting
pair potential.
The eigenfunction expansion of the Green function leads to the usual
representation for the Lorentzian-smoothed DOS
$\dos_{\ipe}(E)$ of the corresponding eigenproblem:
\begin{mathletters}
\bea
\dos_{\ipe}(E)
&\equiv&
\sum_{n}
\frac{1}{\pi}\frac{\ipe}{(E-\qee_{n})^{2}+\ipe^{2}}
\\
&=&
\frac{1}{\pi}
\int_{\pvr}
\lim_{\pvr'\to\pvr}
\tr\,\ima\,\gf(\pvr,\pvr';E+i\ipe),
\label{EQ:DOSExp}
\eea%
\end{mathletters}%
where $\tr$ denotes a trace over e/h components.
\par
We assume that the interface between N and S is a geometrical surface
constituting the boundary of the \ab, i.e., is perfectly sharp.
In other words, $\Delta({\bf r})$ is a constant, $\Delta_{0}$, outside
the billiard and zero inside. Thus, we shall not be working self-consistently,
but shall benefit from being in a position to develop an approach to the
quasiparticle dynamics that focuses on interface-scattering.
\par
To construct an expansion for the Green function $\gf(\pvr,\pvr';z)$
that brings to the fore the geometry of the billiard (i.e.~the spatial
shape of the N-S interface) we adopt the spirit of the Balian-Bloch
approach to the Laplace eigenproblem~\cite{REF:BaBoOne},
and construct a multiple-scattering expansion (MSE)
in which the Green function is represented in terms of the fundamental N
or S Green functions (i.e.~those appropriate for homogeneous N or S
regions).
Although the physical content of this construction is intuitively clear,
its development involves lengthy technical details which we defer to a
forthcoming article~\cite{REF:AGfull}.  The essence of this construction
is the derivation of a system of integral equations \lq\lq residing\rq\rq\
on the N-S interface, the iterative solution of which yields the
aforementioned MSE for the Green function~\cite{REF:BIE}.  Within this
MSE approach, the amplitude for propagating from point $\pvr$ in N to
$\pvr'$ in N, viz.~$\gf(\pvr,\pvr';z)$, is expressed as a sum of the
following processes:
(i)~the \lq\lq free\rq\rq\ propagation amplitude
$\gf^{\rm N}(\pvr,\pvr';z)$;
(ii)~the amplitude involving a single reflection [i.e.~all possible
amplitudes for propagating from $\pvr$ to a generic interface point
$\bbox{\alpha}$, {\it reflecting\/} at $\bbox{\alpha}$,
and then propagating to $\pvr'$:
$-2\int_\alpha \partial\gf^{\rm N}(\pvr,\bbox{\alpha})\,
\psm{3}\,\gf^{\rm N}(\bbox{\alpha},\pvr')$];
(iii)~the amplitude involving two reflections, etc.;
(iv)~the amplitude that traverses the interface twice [i.e.~all
possible amplitudes for propagation from $\pvr$ to the generic
interface point $\bbox{\alpha}$,
{\it transmission\/} into S,
propagation in S from $\bbox{\alpha}$ to another generic interface
point $\bbox{\beta}$,
{\it transmission\/} into N,
and propagation in N from $\bbox{\beta}$ to $\pvr'$:
$-2^2\int_\alpha \partial \gf^{\rm N}(\pvr,\bbox{\alpha})\,\psm{3}\,
\gf^{\rm S}(\bbox{\alpha},\bbox{\beta}) \,
\psm{3}\,\delta\gf^{\rm N}(\bbox{\beta},\pvr')$];
(v)~and so on, where a generic term is specified by an ordered
sequence of reflections and transmissions (see Fig.~\ref{FIG:billiard}d).
Here, $\psm{1,2,3}$ are the Pauli matrices, and
the operators $\partial$ and $\delta$ are defined via
\begin{mathletters}
\begin{eqnarray}
\partial\gf({\pvr},{\bbox{\alpha}})
&\equiv&
{\bf n}_{\alpha}\cdot\bbox{\nabla}_{\psr'}
\gf({\pvr},{\pvr}^{\prime})\vert_{{\pvr}^{\prime}
={\bbox{\alpha}}}\,,
\label{EQ:GFsuperA}
\\
\delta\gf({\bbox{\alpha}},{\pvr}')
&\equiv&
{\bf n}_{\alpha}\cdot\bbox{\nabla}_{\psr\phantom{'}}
\gf({\pvr},{\pvr}^{\prime})\vert_{{\pvr}\phantom{^{\prime}}
={\bbox{\alpha}}}\,,
\label{EQ:GFsuperB}
\end{eqnarray}%
\end{mathletters}%
where ${\bf n}_{\alpha}$ is the normal unit vector pointing
into N at ${\bbox{\alpha}}$ on the N-S interface.
\par\smallskip\noindent
{\it Semiclassical density of states\/}.
So far, our reformulation of the BdG eigenproblem has been exact,
but many of its well-known physical features (such
as the dominance of charge-interconverting reflection processes)
lie hidden beneath the formalism.  They will, however, emerge
when we employ either of two distinct semiclassical (i.e.~short-wave
asymptotic) approximation schemes, as we shall shortly see.
In both schemes, the DOS is calculated via Eq.~(\ref{EQ:DOSExp}), by
using the MSE for $\gf$ and evaluating the resulting integrals using
the stationary-phase approximation, which is appropriate for large
$\Kf L$ and small $\sudel/\Ef$ (where $L$ is the characteristic linear
size of the \ab).  From the technical point of view, the difference
between these schemes lies in the nature of the limits that one assumes
the parameters to take:
\hfil\break\noindent
(A)~$\Kf L\to\infty$ and $\sudel/\Ef\to 0$
with $L\sudel/\Kf$ constant; versus
\hfil\break\noindent
(B)~$\Kf L\to\infty$ with $\sudel/\Ef$ constant.
The limit taken determines which stationary phase points
(i.e.~classical reflection rules) should be applied.
\par
In both schemes, however, it is
possible to integrate out processes involving propagation inside S,
to leading order in $(\Kf L)^{-1}$ and $\sudel/\Ef$.  This is done by
separating each factor of $\gf^{\rm N}$ and $\gf^{\rm S}$ in every kernel
in the MSE into short-ranged pieces and their complements.
By doing this we are distinguishing between {\it local processes\/}
(i.e.~those in which all scatterings take place within a boundary region
of linear size of order $\Kf^{-1}$, so that particles ultimately leave
the boundary region from a point very close to where they first reached it),
and {\it nonlocal processes\/}
(i.e.~the remaining---or long-range---propagation).
Then, we approximate the boundary by the tangent plane at the reflection
point, and evaluate integrals involving short-ranged kernels on this plane.
Moreover, contributions involving the long-ranged part of $\gf^{\rm S}$
are smaller, by a factor of $(\Kf L)^{-1}$, and thus we may neglect
them~\cite{REF:convex}.
This procedure leads to an asymptotic expansion for $\gf$, which
can be used in either of the two semiclassical schemes, and which
includes only interface reflection (as opposed to transmission) and,
correspondingly, involves the renormalized Green
function $\gf^{\rm R}$:
\begin{eqnarray}
&&\gf\simeq
\gf^{\rm N}+
2\negthinspace\int_{\surS}
\negthinspace\partial\gf^{\rm N}\gf^{\rm R}+
2^2 \negthinspace
\int_{\surS}\negthinspace\partial\gf^{\rm N}\,
\partial\gf^{\rm R}\,\gf^{\rm R}+\cdots,
\nonumber
\\
&&\gf^{\rm R}\equiv\Big(
-i{\rm e}^{-i\varphi}
\psm{1}
+{\cal O}
\left(
(\Kf L)^{-2},\sudel/\Ef
\right)\Big)\,\,\gf^{\rm N},
\nonumber
\\
&&\gf^{\rm N}(\pvr,\pvr')\equiv
\pmatrix{g_{+}(\pvr-\pvr') & 0 \cr 0 & -g_{-}(\pvr-\pvr') },
\label{EQ:MRE}
\end{eqnarray}
where $\varphi\equiv\cos^{-1}(E/\sudel)$,
$g_{\pm}(\pvr)\equiv
H_{0}^{\pm}\big(k_{\pm}|{\pvr}|\big)/4$
in two dimensions and
$g_{\pm}(\pvr)\equiv
\exp\left(\pm ik_{\pm}
|{\pvr}|\right)/{4\pi|{\pvr}|}$
in three dimensions,
$k_{\pm}=\sqrt{\Ef\pm E}$ are the e/h wave vectors,
the integrals are taken over the the interface $\surS$,
and, e.g.,
$\int_{\surS}\partial\gf^{\rm N}\gf^{\rm R}\equiv
\int_{\surS}d\bbox{\alpha}\partial
\gf^{\rm N}(\pvr,\bbox{\alpha})\,
\gf^{\rm R}(\bbox{\alpha},\pvr')$.
Observe that the leading term in $\gf^{\rm R}$ includes only
charge-interconverting processes; ordinary reflection appears only
at sub-leading order.  In physical terms, the approximation that
we have invoked takes into account the fact that an electron wave
incident on an N-S interface \lq\lq leaks\rq\rq\ into the S side and,
consequently, is partially converted into a hole and acquires a phase,
much as a particle acquires a phase (i.e.~a Maslov index) when
reflected by a finite single-particle potential.
\par
We are now in a position to define what we shall call the
{\it Perfectly Charge-Interconverting Model\/} (PCIM).  We start
with the expansion~(\ref{EQ:MRE}) for $\gf$ in terms of $\gf^{\rm R}$,
and take the latter to be given by its leading-order form:
$\gf^{\rm R}\approx-i{\rm e}^{-i\varphi}\psm{1}\gf^{\rm N}$.  Then
the PCIM is defined via the following integral equation for $\gf$:
\begin{equation}
\gf=
\gf^{\rm N}
-2i{\rm e}^{-i\varphi}\negthinspace\int_{\surS}
\negthinspace\partial\gf^{\rm N}\psm{1}\gf\,.
\end{equation}
The off-diagonal matrix $\psm{1}$ ensures that, upon each reflection
from the boundary, electrons are fully converted into holes (and
vice versa).  Moreover, this model does retain wave propagation
effects, as implied by the surface integral.
\par
Let us now focus on semiclassical Scheme~A, which is, in spirit,
the one introduced by Andreev~\cite{REF:AFAndreev}.  In this scheme,
excitations undergo perfect retro-reflection (i.e.~perfect
velocity-reversal), as well as perfect charge-interconversion, so
that the dynamics is confined to the geometrical chords of the \ab\
and, thus, is trivially integrable, whatever the shape of the
\ab~\cite{REF:Kosztin1995}.  Via this scheme, we arrive at the
following form for the DOS:
\[
\dos_{\ipe}(E)\simeq
\int_{\surS}\negthinspace\negthinspace
{\rm Re}\,\,
{\cos\theta_{\alpha\beta}\,
 \cos\theta_{\beta\alpha}
 \over{1-
\exp\left[{i(E/\Kf)|{\bbox{\alpha}}
-{\bbox{\beta}}|-2i\varphi}\right]}}
\Big\vert_{E\to E+i\ipe}.
\]
Here, the integral is taken over the surface points
$\bbox{\alpha}$ and $\bbox{\beta}$, and $\theta_{\alpha\beta}$
denotes the angle between the normal at
${\bbox{\alpha}}$ and the chord leading to ${\bbox{\beta}}$.  This
equation for $\dos_{\ipe}$ can be understood as follows: a chord of
length $|{\bbox{\alpha}}-{\bbox{\beta}}|$ contributes eigenvalue weight
at energies given by the well-known semiclassical quantization condition
$E\Kf^{-1}|{\bbox{\alpha}}-{\bbox{\beta}}|-2\cos^{-1}(E/\sudel)=2n\pi$
(for $n$ integral).  However, in order to obtain $\dos_{\ipe}$
we must sum over all chords with the proper weighting, which is
accomplished by the double integral in over the boundary.  The most
prominent features emerging this Scheme~A expression for $\dos_{\ipe}$
are singularities, representing the strong bunching of exact
eigenenergies at energies corresponding to stationary-length chords.
(Such chords have both ends perpendicular to the billiard
boundary.)\thinspace\
However, to sum over {\it all\/} chords would be superfluous, as
the strongest features in the DOS can be captured simply from the
neighborhoods of the stationary chords.  Moreover, for finite
values of the parameters (i.e.~$\Kf L$ large but not infinite,
and $\sudel/\Ef$ small but non-vanishing) Scheme~A produces
a {\it locally averaged\/} DOS, which becomes numerically
accurate only around the DOS singularities that
it predicts.  Thus, it fails to capture this DOS oscillations
due to the confinement of the quasiparticles.  The reason for
this failure is the fact that by summing over all chords one
is implicitly assuming the absence of transverse
quantization/confinement.  To capture such oscillations is the
main motivation for semiclassical Scheme~B.
\par
In Scheme~B we first take into account the imperfectness in
retro-reflection arising from the the previously-neglected
difference between the wave vectors of incident and reflected
electrons and holes, whilst neglecting all amplitudes involving
ordinary reflection.
The corresponding classical dynamics is no longer {\it a priori\/}
integrable; on the contrary, it is chaotic for generic
shapes~\cite{REF:Kosztin1995}.  In this scheme, the closed periodic
orbits fall into two classes, quite distinct from one another:
one consists of multiple
tracings of each stationary chord (we refer to such chords as {\SC}s);
the other of much longer trajectories that \lq\lq creep\rq\rq\
around the billiard boundary (see Fig.~\ref{FIG:billiard}c).
Correspondingly, the DOS is the sum of
(i)~an average term $\dos_{\rm av}(E)$,
which depends in 3D on the volume (or in 2D on the area)
of the billiard (i.e.~the leading Weyl term);
together with an oscillatory term $\dos_{\rm osc}(E)$
consisting of
(ii)~a finer-resolution term, having a universal line-shape
that depends solely on the length and endpoint-curvatures of
the {\SC}s~\cite{REF:isolated}, and
(iii)~very fine resolution terms, which depend on the classical
dynamics of the billiard in question:
\begin{eqnarray}
\dos_{\rm osc}(E)&\simeq&
{\rm Re}\!
\sum_{\ell_{\SCm}}\!
Z_{\SCm}\,{\rm e}^{i\lambda_{\SCm}\pi/4}{\rm Li}_{d-1-\frac{w}{2}}
(1\!-\!{\rm e}^{i(k_{+}-k_{-})\ell_{\SCm}-2i\varphi})
\nonumber \\
&+& \sum_{\rm periodic\ orbits}
A_{\rm po}\,\exp{i S_{\rm po}}\,\,.
\label{EQ:DOSsc}
\end{eqnarray}%
Here,
${\rm Li}_{n}(z)\equiv\sum_{j=1}^{\infty}z^j/j^n$
is the polylogarithm function,
$d$ is the dimensionality of the billiard,
$w$ is the dimensionality of the degeneracy of the \SC\
(e.g.~$w=1$ for a circle),
$Z_{\SCm}$ is a slowly-varying real function of energy,
determining the size of the DOS oscillations, and
$\lambda_{\SCm}$ is a measure of the stability of the \SC,
which determines whether the \lq\lq tail\rq\rq\ goes towards
higher or lower energies.  For example, an {\it isolated\/}
\SC\ in 2D would yield
$\lambda_{\SCm}={\rm sgn}(R_{1}+R_{2}-\ell_{\SCm})-1$ and
\be
Z_{\SCm}=
\sqrt{\frac{(k_{+}+k_{-})^2
\ell_{\SCm}R_{1}R_{2}}{4\pi^2k_{+}k_{-}(k_{+}-
k_{-})^2|\ell_{\SCm}-R_{1}-R_{2}|}},
\nonumber
\ee
where $R_1$ and $R_2$ are the radii of curvature of the endpoints
of the \SC~\cite{REF:Zmode}.
The second term in Eq.~(\ref{EQ:DOSsc}) is the contribution from
\lq\lq creeping\rq\rq\ orbits (see Fig.~\ref{FIG:billiard}c).
In it, $A_{\rm po}$ is determined by the stability of the orbit,
and $S_{\rm po}$ is the action corresponding to the orbit.
For a typical \ab,
$S_{\rm po}>N\big((k_{+}-k_{-})\ell_{\SCm}-2\varphi\big)$, where
$N={\cal O}(\Ef/\sudel)$ and, thus, \lq\lq creeping\rq\rq\ orbits
contribute only to the very fine details of the DOS.
\par
\begin{figure}[hbt]
\epsfxsize=\columnwidth
\centerline{\epsfbox{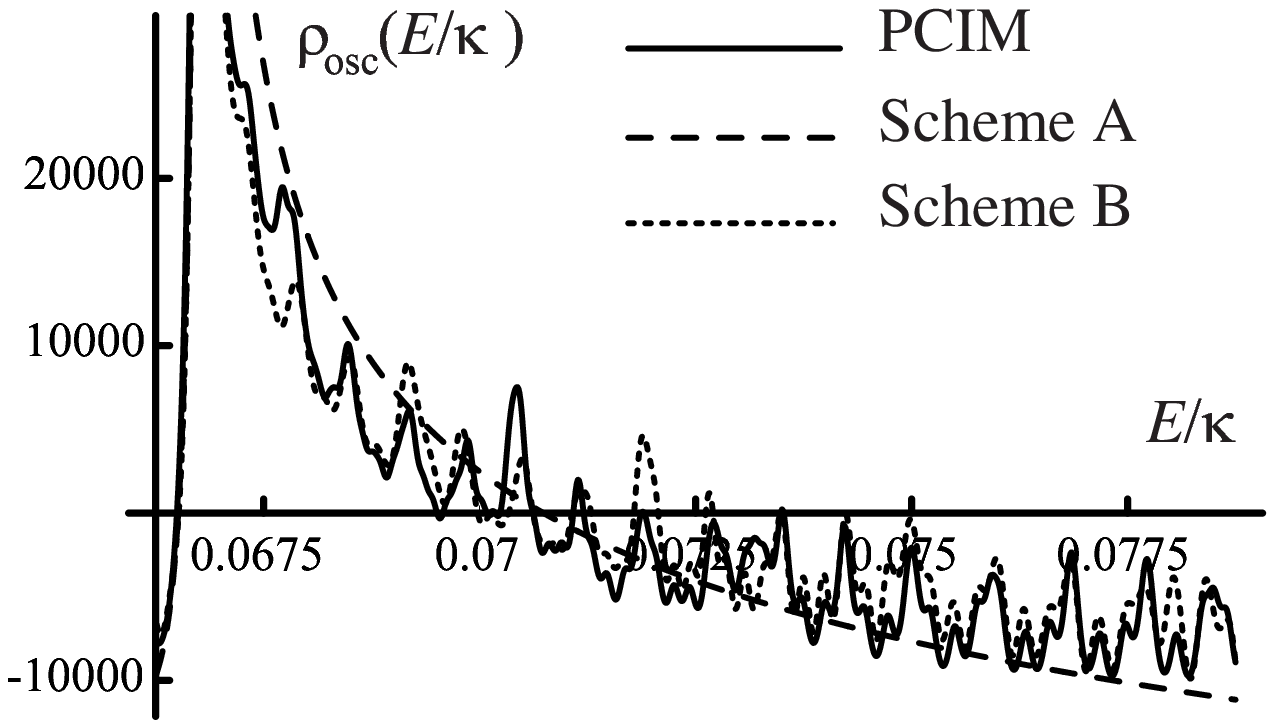}}
\vskip+0.4truecm \caption{Density of states oscillations for a
circular \ab: $\Kf R=150$; $\sudel/\Ef=0.08$; smoothing width
$\Gamma/\Ef=1.1\times 10^{-4}$.}
\label{FIG:CircleDOS1}
\end{figure}%
For illustration, in Fig.~\ref{FIG:CircleDOS1}
we compare the predictions of Schemes~A and B with those of the PCIM.
The Scheme-A result (dashed line) approximates the average behavior
of the exact DOS for the PCIM (full line).  In contrast, the Scheme-B
result (dotted line) captures the DOS oscillations arising from
transverse quantization/confinement.
\par
\begin{figure}[hbt]
\epsfxsize=\columnwidth
\centerline{\epsfbox{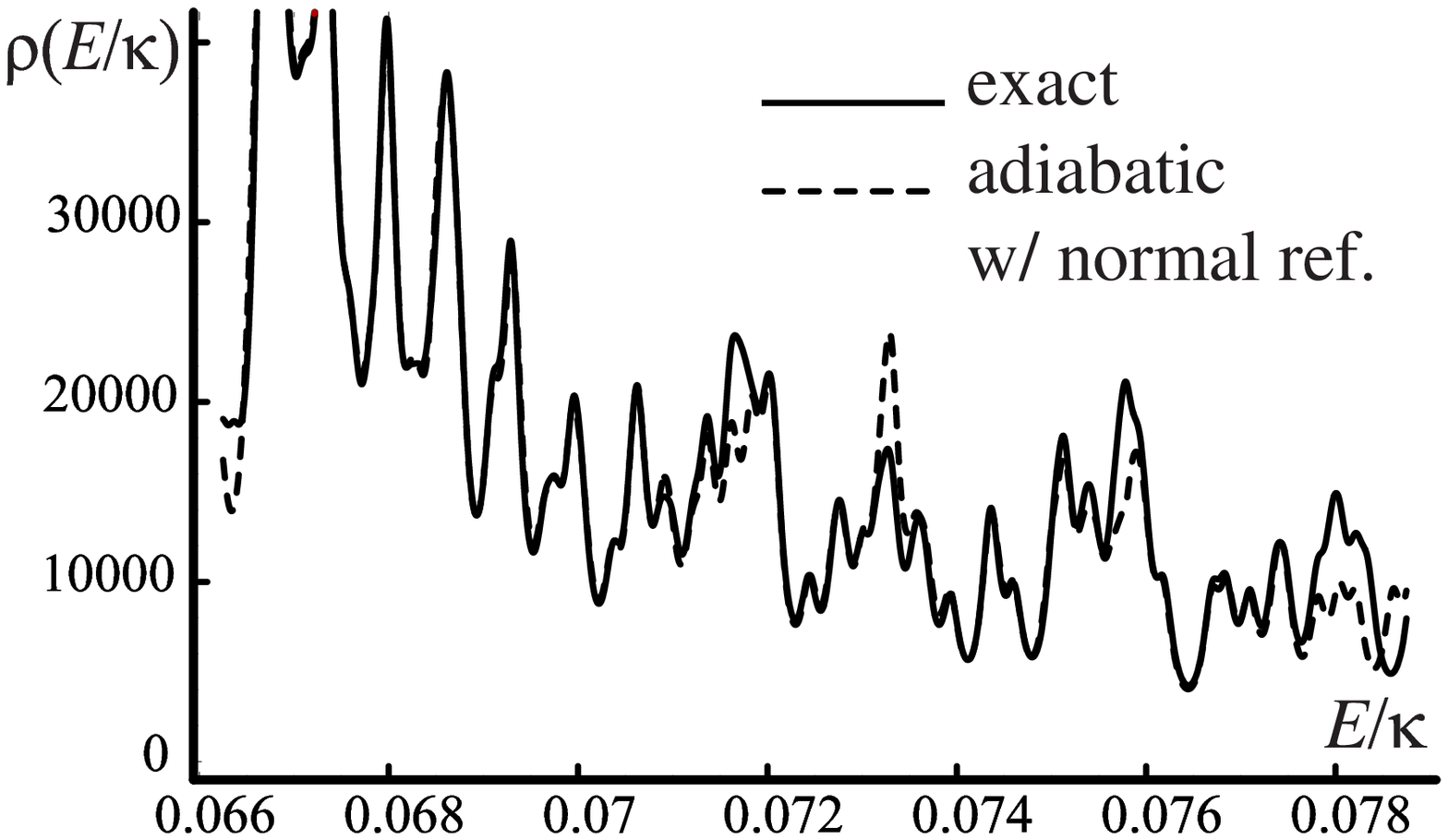}}
\vskip+0.4truecm \caption{Density of states for a
circular \ab: $\Kf R=150$; $\sudel/\Ef=0.08$; smoothing width
$\Gamma/\Ef=1.1\times 10^{-4}$.}
\label{FIG:CircleDOS2}
\end{figure}

Thus far in our semiclassical treatment, we have ignored
all amplitudes involving ordinary reflection.  For non-grazing
incidence [i.e.~$\theta-(\pi/2)\sim 1$] the amplitude for ordinary
reflection is very small ($\sim\sudel/\Ef\cos^2\theta)$).
However, for orbits that contribute dominantly to the oscillatory
structure of the DOS, $\vert\theta-(\pi/2)\vert\ll 1$ and,
therefore, ordinary reflection amplitudes are not negligible and
must be incorporated.  This can be done by
returning to Eq.~(\ref{EQ:MRE}) and re-evaluating the trace formula
using the full expression for $\gf^{\rm R}$ (i.e.~not just the leading,
off-diagonal, term).  However, these dominating orbits are the ones that are
close to the boundary and, for these, consecutive reflections take place
very near to each other, and thus \lq\lq see\rq\rq\ only the local
curvature of the boundary.  These considerations allow us to perform an
\lq\lq adiabatic\rq\rq\ approximation to the expansion in
Eq.~(\ref{EQ:MRE}), in which we assume that the curvature of the boundary
varies slowly, relative to the rate at which creeping orbits sample the
boundary.  In Fig.~\ref{FIG:CircleDOS2} we compare this adiabatic
method with the (exact) result obtained by solving the full BdG
eigenproblem.
\par
We conclude by emphasizing one particular feature of the first term
in Eq.~(\ref{EQ:DOSsc}): this term gives the coarse DOS directly,
through simple geometrical information in the form of the lengths and
endpoint-curvatures of the {\SC}s.  This feature allows the design of
an \ab\ shape that leads to a DOS having a predetermined coarse form.
Moreover, as the stationary-chord terms are well separated
(in time-space) from the creeping orbits, it possible to
\lq\lq hear\rq\rq\ not only the volume of an Andreev
billiard but also its stationary chords.
\par\smallskip\noindent
{\it Acknowledgments\/}.
We gratefully acknowledge useful discussions with
Eric Akkermanns,
Michael Stone and especially
Dmitrii Maslov.
This work was supported by
DOE~DEFG02-96ER45439
and NSF-DMR-99-75187.

\end{multicols}

\begin{references}
\bibitem{REF:Kosztin1995}
Certain classical properties of \ab s
were discussed in
I. Kosztin, D. L. Maslov and P. M. Goldbart,
Phys. Rev. Lett. {\bf 75\/} 1735 (1995).
\bibitem{REF:Rmatrix}
Certain quantum mechanical properties of \ab s
were studied in
A. Altland and M. R. Zirnbauer,
Phys. Rev. Lett. {\bf 76\/}, 3420 (1996);
K. M. Frahm et al.,
Phys. Rev. Lett. {\bf 76\/}, 2981 (1996);
J. A. Melsen et al.,
Europhys. Lett. {\bf 35\/}, 7 (1996);
Physica Scripta T {\bf 69\/}, 223 (1997);
A. Lodder and Yu. V. Nazarov, Phys. Rev. B {\bf 58\/}, 5783 (1998);
H. Schomerus and C. W. J. Beenakker, Phys. Rev. Lett. {\bf 82\/}, 2951 (1999);
W. Ihra et al., cond-mat/9909100.
\bibitem{REF:AFAndreev}
A. F. Andreev,
Zh. Eksp. Teor. Fiz. {\bf 46\/}, 1823 (1964)
[Sov. Phys. J.E.T.P. {\bf 19\/}, 1228 (1964)].
\bibitem{REF:PGdGbook}
See, e.g.,
P.-G. de Gennes,
{\sl Superconductivity of metals and alloys\/}
(Addison-Wesley, New York, 1966), Chap.~5.
\bibitem{REF:BaBoOne}
R. Balian and C. Bloch,
Ann. Phys. (NY) {\bf 60\/}, 401 (1970);
ibid. {\bf 84\/}, 559(E) (1974);
ibid. {\bf 69\/}, 76 (1972).
\bibitem{REF:AGfull}
\.{I}. Adagideli and P. M. Goldbart,
in preparation (2001).
\bibitem{REF:BIE}
For an introduction to boundary integral equation techniques,
see, e.g.,
R. B. Guenther and J. W. Lee,
{\sl Partial differential equations of mathematical physics
and integral equations\/}, (New York: Dover, 1996), Sec.~8-7.
\bibitem{REF:convex}
In fact, for concave shapes there will be nonlocal modifications
that account for tunneling effects.
\bibitem{REF:isolated}
For isolated stationary chords this term has corrections due to changes
in stability that occur when the number of reflections is very large.
\bibitem{REF:Zmode}
The apparent singularity at $E=0$ is an artifact of the
assumption of imperfectness in retro-reflection; this
imperfectness ceases at $E=0$.
\end{references}
\end{document}